\documentclass[12pt,a4paper]{article}
\usepackage{amsmath}
\usepackage{amssymb}
\usepackage{url}
\newcommand{\be}{\begin{equation}}
\newcommand{\ee}{\end{equation}}

\newcommand{\vect}[1]{{\boldsymbol{#1}}}
\newcommand{\vsp}{\vspace*{2mm}}
\newcommand{\vsm}{\vspace*{-2mm}}

\begin{document}
\nopagebreak[4]

\title{MFGSB ({\Large ver. 1.0}): Computer code
  for self-consistent mean-field calculations of atomic nuclei\\
  using Gaussian expansion method\vsp\\
  }
\vspace*{1cm}

\author{H. Nakada\vsp\\
\textit{Department of Physics, Graduate School of Science,
 Chiba University,}\\
\textit{Yayoi-cho 1-33, Inage, Chiba 263-8522, Japan}\vsp\\
\texttt{E-mail: nakada@faculty.chiba-u.jp}}

\date{February 5, 2026}
\maketitle

\begin{abstract}
\texttt{MFGSB}, a computer code
for self-consistent mean-field calculations of atomic nuclei
using the Gaussian expansion method,
is now accessible at Chiba University Repository.
\end{abstract}

\thispagestyle{empty}

\setcounter{page}{0}

\clearpage
\section{Download}\label{ec:download}

The code \texttt{MFGSB} can be downloaded from the following website\,:\\
\centerline{\url{https://opac.ll.chiba-u.jp/da/curator/900123722/?lang=1}}\\
\centerline{(DOI: 10.20776/900123722)}.

The files can be downloaded in two ways,
as described in \texttt{README\_for\_download.txt}.
\begin{enumerate}
\item The zip file \texttt{mfgsb.zip} contains all files,
including the directory structure.
It has 16.6\,GB.
\item Because the interaction data consumes most of the space,
an optional alternative is available
that lets you download the interaction files separately.
The file \texttt{mfgsb\_light.zip} contains the other files.
The individual interaction data are stored
in the folder \texttt{int\_data\_l7}.
When you download the interaction files separately,
it is recommended to create the \texttt{int\_datal7} directory
under the "mfgsb" top directory,
and move the interaction files into it.
\end{enumerate}

\section{Excerpt from UsersGuide\,: Overview}\label{sec:overview}

\texttt{MFGSB} is a code
for the self-consistent mean-field (SCMF) calculations of atomic nuclei,
using the Gaussian expansion method (GEM)
and written in FORTRAN77.
Several special features of the code are described
in Subsec.~\ref{subsec:feature}.

\subsection{System requirements and recommendations}\label{subsec:system}

Memory size depends on the $\ell_\mathrm{cut}$ value.
For the default value $\ell_\mathrm{cut}=7$,
the code spends \underline{8\,GB of memory}.
As the $\ell_\mathrm{cut}$ value increases by one,
about twice the memory will be required.

\underline{The \texttt{BLAS} and \texttt{LAPACK} libraries} should be installed,
which are called in the code.

\subsection{Special features}\label{subsec:feature}

The code has special features owing to the GEM, as below.
\begin{itemize}
\item[(i)\label{adv:int}] Applicable to various functions
  for two-nucleon effective interaction~\cite{ref:NS02}.
  In particular, it is a unique SCMF code
  that is available with the Yukawa interaction up to the tensor channel.
  Also note that the Coulomb exchange term and the two-body term
  of the center-of-mass (c.m.) motion
  can be handled with no additional approximation.
\item[(ii)\label{adv:asym}] Energy-dependent asymptotics
  of the single-particle (s.p.) or quasiparticle (q.p.) wave-functions
  is efficiently described within moderate precision~\cite{ref:NS02,ref:Nak06}.
\item[(iii)\label{adv:basis}] The parameters of the basis functions
  are insensitive to nuclides and to deformation~\cite{ref:Nak08}.
  Thereby a single set of parameters is applicable
  to a wide range of nuclear chart.
  Owing to this feature,
  no need to tune the parameters for the basis functions.
  A file of two-body matrix elements of an interaction
  covers almost all nuclei.
\end{itemize}

\subsection{What can be done?}\label{subsec:what}

Within the SCMF calculations,
the code provides the following opportunities:
\begin{itemize}
\item[0.\label{mf:pot}] Calculations under Woods-Saxon or harmonic oscillator
  potential (not self-consistent);\vsm
\item[1.\label{mf:HF}] Hartree-Fock (HF) calculations;\vsm
\item[2.\label{mf:BCS}] Hartree-Fock plus Bardeen-Cooper-Schrieffer (HF+BCS)
  calculations;\vsm
\item[3.\label{mf:HFB}] Hartree-Fock-Bogolyubov (HFB) calculations;
\end{itemize}
under the following symmetry assumptions,
which are imposed on the one-body fields:
\begin{itemize}
\item[A)\label{sym:sph}] Spherical ($\vect{J}$) symmetry,
  together with parity ($\mathcal{P}$) conservation
  and time-reversal ($\mathcal{T}$) symmetry;\vsm
\item[B)\label{sym:axl}] Axial symmetry
  (rotational symmetry with respect to the $z$-axis, $J_z$),
  together with $\mathcal{P}$ conservation,
  time-reversal ($\mathcal{T}$) symmetry,
  and reflection symmetry with respect to the $y$-axis ($\mathcal{R}$);\vsm
\item[C)\label{sym:apr}] $J_z$ symmetry,
  together with $\mathcal{P}$ conservation
  and $\mathcal{RT}$ symmetry (product of $\mathcal{R}$ and $\mathcal{T}$).
\end{itemize}
Treatment of the c.m. motion is selected by option.
To B) and C),
the constraining term of the quadrupole moment can be added.

Regarding numerical methods to attain self-consistency,
either the iterative method or the conjugate gradient method can be used.
(For HF+BCS, the method can be chosen for the HF part,
while the iterative method is applied to the BCS part.)
In the iterative method,
the HF or HFB Hamiltonian is diagonalized every time.
In the conjugate gradient method,
the HF or HFB Hamiltonian is diagonalized at the end.

\subsection{Terms of use}\label{subsec:terms}

For copyright, see \texttt{LICENSE.txt}.
When publishing results using this code,
cite the following references,
in which the numerical methods were developed and established~
~\cite{ref:NS02,ref:Nak06,ref:Nak08,ref:NI25}.


\end{document}